\documentclass[prb,aps,showpacs,twocolumn]{revtex4}
\usepackage{amssymb}
\usepackage{amsmath}
\usepackage{graphicx}

\input{tcilatex}

\begin{document}

\title{\textbf{Quantum states transfer by the analogous Bell states}}
\author{Di Mei}
\author{Chong Li }
\author{Guo-Hui Yang}
\author{He-Shan Song}\email{hssong@dlut.edu.cn}
\affiliation{Department of Physics, Dalian University of
Techology, Dalian 116024, P. R. China }

\begin{abstract}
Transmitting quantum states by channels of analogous Bell states
is studied in this paper. We analyse the transmitting process,
constructed the probabilitic unitary operator, and gain the
largest successful transfer quantum state probability.
\end{abstract}

\maketitle

\section{Introduction}

It is well known that there are two ways to transfer quantum
information: one can first use the channel to share entanglement
with separated Alice and Bob and then use this entanglement for
teleportation [1], or directly transmit a state through a quantum
data bus. In the quantum information theory [2], transfer of
information in the form of a prepared superposition quantum state
is essential. And the important task in quantum-information
processing is the transfer of quantum states from one location A
to another location B[3]. One can transfer a quantum state either
by the method of teleportation [1] or through quantum networking.
The typical examples of quantum state transfer is the quantum
storage based on various physical systems [4,5], such as the
quasispin wave excitations [6]. Recently, schemes have been
proposed that employ more than two qubits to perform various
quantum information tasks [7,8,9]. There is also an interest in
performing quantum teleportation of state of more than one qubit.
Lee[10] has presented a setup for teleportation of an entangled
state of two photons, such scheme as some other schemes [11], uses
photons because they propagate fast and can carry quantum
information over long distances. A state transfer between two
identical distant systems is a process in which at time t=T the
second system obtains the same quantum state that the first one
had at time t=0[12]. In Ref.[13], it referces to the transfer
quantum states by Bell states. We know that if the entanglement
pairs are Bell states, the probability of transmitting quantum
states is $100\%$. Generally, the Bell states are prepared via
evolvement of H till $t=t_{0}$ which is the time that the states
are prepared as Bell states. For example
\begin{equation*}
\left\vert \phi \right\rangle =\sin t\left\vert 00\right\rangle
+\cos t\left\vert 11\right\rangle
\end{equation*}%
when $t=\frac{\pi }{4},$%
\begin{equation*}
\left\vert \phi \right\rangle =\frac{1}{\sqrt{2}}\left\vert 00\right\rangle +%
\frac{1}{\sqrt{2}}\left\vert 11\right\rangle .
\end{equation*}%
But for the difficulty of exactly controling the process, it is
nearly no possibility to prepare the perfect Bell states. It is
easy to produce some departures from Bell states. It is to say
\begin{equation*}
\left\vert \phi \right\rangle =\sin (\frac{\pi }{4}+\triangle
)\left\vert 00\right\rangle +\cos (\frac{\pi }{4}+\triangle
)\left\vert 11\right\rangle .
\end{equation*}%
In this paper, we name these states as the analogous Bell states.
This condition will lead to be difficult to transmit quantum
states of the system. Therefor, we try to deal with these troubles
and specially we use one qubit of entanglement state in each
entanglement pair as the channels to transmission.

In the Sec.II and Sec.III, we accomplish the probability of
transmission quantum states by these analogous Bell states and
some effective unitary operators. And we exhibit the relation
between the probability and the departures with Fig.1 and Fig.2.

\section{Transmitting bipartite quantum states by two analogous Bell states}

Here we suppose that Alice and Bob share two pairs entanglement
particles. We suppose the entanglement states are:

\begin{eqnarray}
\left\vert \phi \right\rangle _{A} &=&a_{A}\left\vert
00\right\rangle _{A}+b_{A}\left\vert 11\right\rangle _{A}\text{ \
\ }\left\vert
a_{A}\right\vert ^{2}+\left\vert b_{A}\right\vert ^{2}=1 \\
&&  \notag \\
a_{A} &=&\sin (\frac{\pi }{4}+\triangle _{A})\text{ \ \ }b_{A}=\cos (\frac{%
\pi }{4}+\triangle _{A})  \notag \\
&&  \notag \\
\left\vert \phi \right\rangle _{B} &=&a_{B}\left\vert
01\right\rangle _{B}+b_{B}\left\vert 10\right\rangle _{B}\text{ \
\ }\left\vert
a_{B}\right\vert ^{2}+\left\vert b_{B}\right\vert ^{2}=1 \\
&&  \notag \\
a_{B} &=&\sin (\frac{\pi }{4}+\triangle _{B})\text{ \ \ }b_{B}=\cos (\frac{%
\pi }{4}+\triangle _{B}).  \notag
\end{eqnarray}

We show how to transmit an arbitrary bipartite state by two
analogous Bell states

\begin{equation}
\left\vert \phi \right\rangle _{12}=c_{1}\left\vert
00\right\rangle _{12}+c_{2}\left\vert 01\right\rangle
_{12}+c_{3}\left\vert 10\right\rangle _{12}+c_{4}\left\vert
11\right\rangle _{12}
\end{equation}

The whole state is
\begin{eqnarray}
\left\vert \Phi \right\rangle &=&a_{A}a_{B}c_{1}\left\vert
00\right\rangle _{12}\left\vert 00\right\rangle _{A\text{
}}\left\vert 01\right\rangle _{B}+a_{A}b_{B}c_{2}\left\vert
01\right\rangle _{12}\left\vert
00\right\rangle _{A}\left\vert 10\right\rangle _{B}  \notag \\
&&+b_{A}a_{B}c_{3}\left\vert 10\right\rangle _{12}\left\vert
11\right\rangle _{A}\left\vert 01\right\rangle
_{B}+b_{A}b_{B}c_{4}\left\vert 11\right\rangle _{12}\left\vert
11\right\rangle _{A}\left\vert
10\right\rangle _{B}.  \notag \\
&&
\end{eqnarray}

Then by carrying out Bell-operation on the first particle, we get%
\begin{eqnarray}
\left\vert \Phi
{\acute{}}%
\right\rangle &=&\frac{1}{\sqrt{2}}(a_{A}a_{B}c_{1}\left\vert
00\right\rangle _{12}\left\vert 00\right\rangle _{A}\left\vert
01\right\rangle _{B}+  \notag \\
&&a_{A}a_{B}c_{1}\left\vert 10\right\rangle _{12}\left\vert
00\right\rangle _{A}\left\vert 01\right\rangle
_{B}+a_{A}b_{B}c_{2}\left\vert 01\right\rangle _{12}\left\vert
00\right\rangle _{A}\left\vert
10\right\rangle _{B}  \notag \\
&&+a_{A}b_{B}c_{2}\left\vert 11\right\rangle _{12}\left\vert
00\right\rangle _{A}\left\vert 10\right\rangle
_{B}+b_{A}a_{B}c_{3}\left\vert 00\right\rangle _{12}\left\vert
11\right\rangle _{A}\left\vert
01\right\rangle _{B}  \notag \\
&&-b_{A}a_{B}c_{3}\left\vert 10\right\rangle _{12}\left\vert
11\right\rangle _{A}\left\vert 01\right\rangle
_{B}+b_{A}b_{B}c_{4}\left\vert 01\right\rangle _{12}\left\vert
11\right\rangle _{A}\left\vert
10\right\rangle _{B}  \notag \\
&&-b_{A}b_{B}c_{4}\left\vert 11\right\rangle _{12}\left\vert
11\right\rangle _{A}\left\vert 10\right\rangle _{B})
\end{eqnarray}

The measurement operator $M$ is
\begin{eqnarray}
M &=&c(\left\vert 000000\right\rangle \left\langle
000000\right\vert
+\left\vert 000001\right\rangle \left\langle 000001\right\vert  \notag \\
&&+\left\vert 000010\right\rangle \left\langle 000010\right\vert
+\left\vert
000100\right\rangle \left\langle 000100\right\vert  \notag \\
&&+\cdots +\left\vert 001111\right\rangle \left\langle
001111\right\vert
\notag \\
&&+\cdots +\left\vert 011111\right\rangle \left\langle
011111\right\vert )
\end{eqnarray}%
$c$ is a normalization constant.

After measurement, we got%
\begin{eqnarray}
M\left\vert \Phi
{\acute{}}%
\right\rangle &=&a_{A}a_{B}c_{1}\left\vert 00\right\rangle
\left\vert
00\right\rangle \left\vert 01\right\rangle +  \notag \\
&&a_{A}b_{B}c_{2}\left\vert 01\right\rangle \left\vert
00\right\rangle \left\vert 10\right\rangle
+b_{A}a_{B}c_{3}\left\vert 00\right\rangle
\left\vert 11\right\rangle \left\vert 00\right\rangle  \notag \\
&&+b_{A}b_{B}c_{4}\left\vert 01\right\rangle \left\vert
11\right\rangle \left\vert 10\right\rangle
\end{eqnarray}

To realize the transmission, we require two unitary operators. One
is $U$
which is expressed by a $2^{6}\times 2^{6}$ matrix.%
\begin{eqnarray*}
U_{1,2} &=&1\text{ \ }U_{2,1}=1\text{ \ }U_{1,1}=0\text{ \ }U_{2,2}=0 \\
U_{3,19} &=&1\text{ \ }U_{19,3}=1\text{ }U_{3,3}=0\text{ \ }U_{19,19}=0 \\
U_{5,14} &=&1\text{ \ }U_{14,5}=1\text{ \ }U_{5,5}=0\text{ \ }U_{14,14}=0 \\
U_{7,31} &=&1\text{ \ }U_{31,7}=1\text{ \ }U_{7,7}=0\text{ \ }U_{31,31}=0 \\
others\text{ }U_{i,i} &=&1\text{ \ }and\text{ \ }U_{i,j}=0\text{ \
}i\neq j
\end{eqnarray*}%
then the state is
\begin{eqnarray}
\left\vert \Psi \right\rangle &=&m_{1}c_{1}\left\vert
00\right\rangle \left\vert 00\right\rangle \left\vert
00\right\rangle +m_{2}c_{2}\left\vert 00\right\rangle \left\vert
00\right\rangle \left\vert 10\right\rangle  \notag
\\
&&+m_{3}c_{3}\left\vert 00\right\rangle \left\vert 01\right\rangle
\left\vert 00\right\rangle +m_{4}c_{4}\left\vert 00\right\rangle
\left\vert 01\right\rangle \left\vert 10\right\rangle
\end{eqnarray}%
where $m_{1}=a_{A}a_{B},$ $m_{2}=a_{A}b_{B},$ $m_{3}=b_{A}a_{B},$ $%
m_{4}=b_{A}b_{B}$ and we perform the second operator $F$ that is also a $%
2^{6}\times 2^{6}$ matrix.
\begin{eqnarray*}
F_{3,3} &=&\frac{m_{1}}{m_{2}}\text{ \ }F_{5,5}=\frac{m_{1}}{m_{3}}\text{ \ }%
F_{7,7}=\frac{m_{1}}{m_{4}} \\
\text{ \ }F_{8,8} &=&0\text{ \ \ \ \ }F_{9,9}=0 \\
\text{ \ \ \ \ }F_{10,10} &=&0\text{ \ \ \ \ }F_{11,11}=0 \\
F_{3,9} &=&\sqrt{1-\left\vert \frac{m_{1}}{m_{2}}\right\vert
^{2}}\text{ \ \
\ }F_{8,3}=-\sqrt{1-\left\vert \frac{m_{1}}{m_{2}}\right\vert ^{2}} \\
F_{8,9} &=&\left( \frac{m_{1}}{m_{2}}\right) ^{\ast } \\
F_{5,10} &=&\sqrt{1-\left\vert \frac{m_{1}}{m_{3}}\right\vert
^{2}}\text{ \
\ \ }F_{9,5}=-\sqrt{1-\left\vert \frac{m_{1}}{m_{3}}\right\vert ^{2}} \\
F_{9,10} &=&\left( \frac{m_{1}}{m_{3}}\right) ^{\ast } \\
F_{7,11} &=&\sqrt{1-\left\vert \frac{m_{1}}{m_{4}}\right\vert
^{2}}\text{ \
\ \ }F_{10,7}=-\sqrt{1-\left\vert \frac{m_{1}}{m_{4}}\right\vert ^{2}} \\
F_{10,11} &=&\left( \frac{m_{1}}{m_{4}}\right) ^{\ast }\text{ \ \ \ \ \ \ \ }%
F_{11,8}=1 \\
other\text{ }F_{i,i} &=&1,\text{ }F_{i,j}=0\text{\ \ \ }
\end{eqnarray*}%
here we suppose $m_{1}$ is the least. We get the state%
\begin{eqnarray}
\left\vert \Psi ^{\prime }\right\rangle &=&m_{1}(c_{1}\left\vert
00\right\rangle \left\vert 00\right\rangle \left\vert
00\right\rangle +c_{2}\left\vert 00\right\rangle \left\vert
00\right\rangle \left\vert
10\right\rangle +  \notag \\
&&c_{3}\left\vert 00\right\rangle \left\vert 01\right\rangle
\left\vert 00\right\rangle +c_{4}\left\vert 00\right\rangle
\left\vert 01\right\rangle \left\vert 10\right\rangle )
\end{eqnarray}%
we perform the projective measurement given as

\begin{equation}
P_{s}=\left\vert 00\right\rangle _{12}\left\vert 0\right\rangle
_{A_{1}}\left\vert 0\right\rangle _{B_{2}B_{2}}\left\langle
0\right\vert _{A_{1}}\left\langle 0\right\vert _{12}\left\langle
00\right\vert
\end{equation}%
if we get 1, it shows the success of us. The probability is easily got as $%
4\left\vert m_{1}\right\vert ^{2}$ which is equal to $4\left\vert \sin (%
\frac{\pi }{4}+\triangle _{A})\sin (\frac{\pi }{4}+\triangle
_{B})\right\vert ^{2}.$ For more distinctly to observe, we propose
the FIG.1 to exhibit the relation between the probability and
$\triangle _{A},\triangle _{B}.$

\section{transmit quantum states of multipartite particles by multipartite
analogous Bell states}

In this section, we try to apply our scheme to the multipartite
particles system. So we take the transfer of three particles
quantum states by three analogous Bell states as the example. We
have three pairs of entanglement paticles whose states are

\begin{eqnarray}
\left\vert \phi \right\rangle _{A} &=&a_{A}\left\vert
00\right\rangle
_{A}+b_{A}\left\vert 11\right\rangle _{A} \\
&&  \notag \\
\left\vert \phi \right\rangle _{B} &=&a_{B}\left\vert
01\right\rangle
_{B}+b_{B}\left\vert 10\right\rangle _{B} \\
&&  \notag \\
\left\vert \phi \right\rangle _{C} &=&a_{C}\left\vert
00\right\rangle _{C}-b_{C}\left\vert 11\right\rangle _{C}\text{ }
\end{eqnarray}

The state of the syetem of three particles is

\begin{eqnarray}
\left\vert \phi \right\rangle _{123} &=&c_{1}\left\vert
000\right\rangle _{123}+c_{2}\left\vert 001\right\rangle
_{123}+c_{3}\left\vert
010\right\rangle _{123}  \notag \\
&&+c_{4}\left\vert 011\right\rangle _{123}+c_{5}\left\vert
100\right\rangle
_{123}+c_{6}\left\vert 101\right\rangle _{123}  \notag \\
&&+c_{7}\left\vert 110\right\rangle _{123}+c_{8}\left\vert
111\right\rangle _{123}
\end{eqnarray}

Therefor, we gain the whole state which is

\begin{eqnarray}
\left\vert \Phi \right\rangle &=&\left\vert \phi \right\rangle
_{123}\left\vert \phi \right\rangle _{A}\left\vert \phi
\right\rangle
_{B}\left\vert \phi \right\rangle _{C}  \notag \\
&=&c_{1}a_{A}a_{B}a_{C}\left\vert 000\right\rangle
_{123}\left\vert 00\right\rangle _{A\text{ }}\left\vert
01\right\rangle _{B}\left\vert
00\right\rangle _{C}+\cdots -  \notag \\
&&c_{2}a_{A}a_{B}b_{C}\left\vert 001\right\rangle _{123}\left\vert
00\right\rangle _{A\text{ }}\left\vert 01\right\rangle
_{B}\left\vert
11\right\rangle _{C}+\cdots +  \notag \\
&&c_{3}a_{A}b_{B}b_{C}\left\vert 010\right\rangle _{123}\left\vert
00\right\rangle _{A\text{ }}\left\vert 10\right\rangle
_{B}\left\vert
00\right\rangle _{C}+\cdots -  \notag \\
&&c_{4}a_{A}b_{B}b_{C}\left\vert 011\right\rangle _{123}\left\vert
00\right\rangle _{A\text{ }}\left\vert 10\right\rangle
_{B}\left\vert
11\right\rangle _{C}+\cdots +  \notag \\
&&c_{5}b_{A}a_{B}a_{C}\left\vert 100\right\rangle _{123}\left\vert
11\right\rangle _{A\text{ }}\left\vert 01\right\rangle
_{B}\left\vert
00\right\rangle _{C}+\cdots -  \notag \\
&&c_{6}b_{A}a_{B}b_{C}\left\vert 101\right\rangle _{123}\left\vert
11\right\rangle _{A\text{ }}\left\vert 01\right\rangle
_{B}\left\vert
11\right\rangle _{C}+\cdots +  \notag \\
&&c_{7}b_{A}b_{B}a_{C}\left\vert 110\right\rangle _{123}\left\vert
11\right\rangle _{A\text{ }}\left\vert 10\right\rangle
_{B}\left\vert
00\right\rangle _{C}+\cdots -  \notag \\
&&c_{8}b_{A}b_{B}b_{C}\left\vert 111\right\rangle _{123}\left\vert
11\right\rangle _{A\text{ }}\left\vert 10\right\rangle
_{B}\left\vert 11\right\rangle _{C}+\cdots
\end{eqnarray}

To prepare the state to the expected state, we propose a unitary
operator $U$ which is a $2^{9}\times 2^{9}$ matrix. The elements
of $U$ are

\begin{eqnarray*}
U_{7,72} &=&1\text{ \ }U_{72,7}=1\text{ \ }U_{7,7}=0\text{ \ }U_{72,72}=0 \\
U_{13,137} &=&1\text{ \ }U_{137,13}=1\text{ }U_{13,13}=0\text{ \ }%
U_{137,137}=0 \\
U_{15,204} &=&1\text{ \ }U_{204,15}=1\text{ \ }U_{15,15}=0\text{ \ }%
U_{204,204}=0 \\
U_{37,307} &=&1\text{ \ }U_{307,37}=1\text{ \ }U_{37,37}=0\text{ \ }%
U_{307,307}=0 \\
U_{39,366} &=&1\text{ \ }U_{366,39}=1\text{ \ }U_{39,39}=0\text{ \ }%
U_{366,366}=0 \\
U_{45,441} &=&1\text{ \ }U_{441,45}=1\text{ \ }U_{45,45}=0\text{ \ }%
U_{441,441}=0 \\
U_{47,508} &=&1\text{ \ }U_{508,47}=1\text{ \ }U_{47,47}=0\text{ \ }%
U_{508,508}=0 \\
others\text{ }U_{i,i} &=&1\text{ \ }and\text{ \ }U_{i,j}=0\text{ \
}i\neq j
\end{eqnarray*}

Then, we find that the coefficients of the terms are different.
This will lead to the condition that we could not get the ideal
results. To deal with it, we propose the second unitary which is
$F.$ It is also a $2^{9}\times 2^{9}$ matrix. And the elements are

\begin{eqnarray*}
F_{7,7} &=&\frac{m_{1}}{m_{2}}\text{ \
}F_{13,13}=\frac{m_{1}}{m_{3}}\text{
\ }F_{17,17}=\frac{m_{1}}{m_{4}} \\
F_{37,37} &=&\frac{m_{1}}{m_{5}}\text{ \
}F_{39,39}=\frac{m_{1}}{m_{6}}\text{
\ }F_{45,45}=\frac{m_{1}}{m_{7}} \\
F_{47,47} &=&\frac{m_{1}}{m_{8}} \\
F_{7,65} &=&\sqrt{1-\left\vert \frac{m_{1}}{m_{2}}\right\vert
^{2}}\text{ \
\ \ }F_{64,7}=-\sqrt{1-\left\vert \frac{m_{1}}{m_{2}}\right\vert ^{2}} \\
F_{64,65} &=&\left( \frac{m_{1}}{m_{2}}\right) ^{\ast } \\
F_{13,66} &=&\sqrt{1-\left\vert \frac{m_{1}}{m_{3}}\right\vert
^{2}}\text{ \
\ \ }F_{65,13}=-\sqrt{1-\left\vert \frac{m_{1}}{m_{3}}\right\vert ^{2}} \\
F_{65,66} &=&\left( \frac{m_{1}}{m_{3}}\right) ^{\ast } \\
F_{15,67} &=&\sqrt{1-\left\vert \frac{m_{1}}{m_{4}}\right\vert
^{2}}\text{ \
\ \ }F_{66,15}=-\sqrt{1-\left\vert \frac{m_{1}}{m_{4}}\right\vert ^{2}} \\
F_{66,67} &=&\left( \frac{m_{1}}{m_{4}}\right) ^{\ast }\text{ \ \
\ \ \ \ \ }
\\
F_{37,68} &=&\sqrt{1-\left\vert \frac{m_{1}}{m_{5}}\right\vert
^{2}}\text{ \
\ \ }F_{67,37}=-\sqrt{1-\left\vert \frac{m_{1}}{m_{5}}\right\vert ^{2}} \\
F_{67,68} &=&\left( \frac{m_{1}}{m_{5}}\right) ^{\ast } \\
F_{39,69} &=&\sqrt{1-\left\vert \frac{m_{1}}{m_{6}}\right\vert
^{2}}\text{ \
\ \ }F_{68,39}=-\sqrt{1-\left\vert \frac{m_{1}}{m_{6}}\right\vert ^{2}} \\
F_{68,69} &=&\left( \frac{m_{1}}{m_{6}}\right) ^{\ast } \\
F_{45,70} &=&\sqrt{1-\left\vert \frac{m_{1}}{m_{7}}\right\vert
^{2}}\text{ \
\ \ }F_{69,45}=-\sqrt{1-\left\vert \frac{m_{1}}{m_{7}}\right\vert ^{2}} \\
F_{69,70} &=&\left( \frac{m_{1}}{m_{7}}\right) ^{\ast } \\
F_{47,71} &=&\sqrt{1-\left\vert \frac{m_{1}}{m_{8}}\right\vert
^{2}}\text{ \
\ \ }F_{70,47}=-\sqrt{1-\left\vert \frac{m_{1}}{m_{8}}\right\vert ^{2}} \\
F_{70,71} &=&\left( \frac{m_{1}}{m_{8}}\right) ^{\ast } \\
F_{64,64} &=&F_{65,65}=F_{66,66}=F_{67,67}=F_{68,68}=F_{69,69} \\
&=&F_{70,70}=F_{71,71}=0 \\
F_{71,64} &=&1\text{ \ \ \ \ }other\text{ }F_{i,i}=1,\text{ }F_{i,j}=0\text{%
\ \ \ }
\end{eqnarray*}%
where $m_{1}=a_{A}a_{B}a_{C},$ $m_{2}=a_{A}a_{B}b_{C},$ $%
m_{3}=a_{A}b_{B}b_{C},$ $m_{4}=a_{A}b_{B}b_{C},$ $m_{5}=b_{A}a_{B}a_{C},$ $%
m_{6}=b_{A}a_{B}b_{C},$ $m_{7}=b_{A}b_{B}a_{C},$
$m_{8}=b_{A}b_{B}b_{C}.$ And we suppose the $\left\vert
m_{1}\right\vert $ is the least of all the coefficients of $F.$

After measurement, the state is

\begin{eqnarray}
\left\vert \Psi \right\rangle &=&m_{1}(c_{1}\left\vert
000\right\rangle _{123}\left\vert 00\right\rangle _{A\text{
}}\left\vert 01\right\rangle
_{B}\left\vert 00\right\rangle _{C}+  \notag \\
&&c_{2}\left\vert 000\right\rangle _{123}\left\vert
00\right\rangle _{A\text{
}}\left\vert 01\right\rangle _{B}\left\vert 10\right\rangle _{C}+  \notag \\
&&c_{3}\left\vert 000\right\rangle _{123}\left\vert
00\right\rangle _{A\text{
}}\left\vert 11\right\rangle _{B}\left\vert 00\right\rangle _{C}+  \notag \\
&&c_{4}\left\vert 000\right\rangle _{123}\left\vert
00\right\rangle _{A\text{
}}\left\vert 11\right\rangle _{B}\left\vert 10\right\rangle _{C}+  \notag \\
&&c_{5}\left\vert 000\right\rangle _{123}\left\vert
10\right\rangle _{A\text{
}}\left\vert 01\right\rangle _{B}\left\vert 00\right\rangle _{C}+  \notag \\
&&c_{6}\left\vert 000\right\rangle _{123}\left\vert
10\right\rangle _{A\text{
}}\left\vert 01\right\rangle _{B}\left\vert 10\right\rangle _{C}+  \notag \\
&&c_{7}\left\vert 000\right\rangle _{123}\left\vert
10\right\rangle _{A\text{
}}\left\vert 11\right\rangle _{B}\left\vert 00\right\rangle _{C}+  \notag \\
&&c_{8}\left\vert 000\right\rangle _{123}\left\vert
10\right\rangle _{A\text{
}}\left\vert 11\right\rangle _{B}\left\vert 10\right\rangle _{C})+  \notag \\
&&other\text{ }terms
\end{eqnarray}

Observing the state, it is easy to find that we could gain the
quantum state of transfer by projective measurement $P_{s}$ which
is

\begin{equation}
P_{s}=\left\vert 000\right\rangle _{123}\left\vert 0\right\rangle
_{A_{2}}\left\vert 1\right\rangle _{B_{2}}\left\vert
0\right\rangle _{C_{2}C_{2}}\left\langle 0\right\vert
_{B_{2}}\left\langle 1\right\vert _{A_{2}}\left\langle
0\right\vert _{123}\left\langle 000\right\vert
\end{equation}

Therefor, we apply the scheme of transfer quantum states of
multipartite particles by multipartite analogous Bell states
successfully.

\section{conclusion}

It shows that we could be successful to complete the transfer of
bipartite quantum states and quantum states of multipartite
particles by the analogous Bell states, and the unitary operators
are effective to overcome the difficulties that are produced by
the departures. From Fig.1 and Fig.2, we could distinctly know the
relation between the probability transmission and the departures,
For example, when the departures are equal to 0 that it means the
channels are the Bell states, the probability is 1. Therefor we
achieve our tasks successfully.

\begin{acknowledgments}
This work was partially supported by the CNSF (grant No.10575017)
.
\end{acknowledgments}

\end{document}